\documentclass{WileyMSP-template}
\usepackage{textcomp}
\usepackage{ragged2e}
\usepackage{parskip}
\usepackage{amsmath}
\begin{document}

\pagestyle{fancy}
% \rhead{\includegraphics[width=2.5cm]{vch-logo.png}}

\vspace{\baselineskip}
\vspace{\baselineskip}

\title{Thermally Drawn Bioelectric Catheters: Enabling Proprioceptive Endovascular Navigation}

\maketitle

\vspace{\baselineskip}

\author{Alex Ranne*,}
\author{Heiko Maier*,}
\author{Jinshi Zhao*,}
\author{Songli Moey,}
\author{Ayhan Aktas,}
\author{Manik Chana,}
\author{Burak Temelkuran,}
\author{Nassir Navab,}
\author{Ferdinando Rodriguez y Baena\textsuperscript{\textdagger}}\\

\vspace{\baselineskip}
\vspace{\baselineskip}

$^*$ These authors contributed equally to this work.\\
\textsuperscript{\textdagger} Corresponding author: Ferdinando Rodriguez y Baena, Email: f.rodriguez@imperial.ac.uk \\
\begin{affiliations}
A. Ranne*, S. L. Moey, A. Aktas, F. R. y Baena \\
\emph{The Hamlyn Centre for Robotic Surgery, Imperial College London, SW7 2AZ, London, UK}\\

H. Maier*, N. Navab\\
\emph{Chair for Computer Aided Medical Procedures and Augmented Reality (CAMP), Technical University of Munich, 85748 Garching, Germany }\\

J. S. Zhao*, B. Temelkuran\\
\emph{Department of Metabolism, Digestion and Reproduction, Imperial College London, SW7 2AZ, London, UK} \\

M. Chana \\
\emph{Major Trauma Centre, St. Mary's Hospital, W2 1NY, London, UK}\\

\end{affiliations}
\vspace{\baselineskip}

\keywords{Cardiovascular Disease, Personalized Medicine, Thermal Drawing, Biomedical Engineering, Impedance Sensing}

\vspace{\baselineskip}

\begin{justify}
\begin{abstract}

To navigate medical instruments safely and accurately inside a patient’s vascular tree, combining X-ray fluoroscopy with intermittent contrast injections is the gold standard. However, prolonged exposure to ionizing radiation poses health risks, necessitates the use of cumbersome lead vests for the clinicians, and contrast injections can lead to acute kidney injury in patients. Bioelectric Navigation, a non-fluoroscopic tracking modality, aims to provide an alternative. It uses weak electric currents to detect local anatomical features in the vasculature and localize instruments without x-ray imaging. 
In this work, we advance Bioelectric Navigation on two frontiers. Firstly, we introduce a new class of bespokely designed electrode catheters. They are fabricated using 3D printing, thermal drawing, and laser micro-machining. Specifically, we manufacture a 6 Fr catheter incorporating 16 electrodes, a guidewire channel and an additional sensor compartment. We thoroughly assess the catheter’s mechanical and electrical properties. Secondly, we introduce an algorithm to localize the catheter along the centerline of a vascular phantom, for the first time fusing electric detection of vascular geometry with electric distance estimation. We report both tracking accuracy and usability evaluated by an expert endovascular surgeon, demonstrating the strong potential of this technology for integration into the existing clinical workflow.

\end{abstract}

\vspace{\baselineskip}

\section{Introduction}

Cardiovascular diseases (CVD) are the most common cause of death worldwide \cite{kaplan2016kaplan}. Many CVD can be treated interventionally, with minimally invasive endovascular surgery (MIES) being the gold standard. In MIES procedures, surgeons insert long, thin, tubular devices into the patient's vascular system, such as catheters and guidewires. These devices are advanced through the blood vessels to the diseased region under fluoroscopic guidance. Endovascular aneurysm repair (EVAR) is a type of MIES that treats weakened, bulging vessel walls (aneurysms) by introducing stabilizing stent grafts. Other procedures include the treatment of stenotically narrowed blood vessels that restrict blood flow to critical organs like the heart muscle or brain, through angioplasty and stenting \cite{kessel2016interventional,beard2013vascular}. In all these procedures, having a reliable form of device tracking and navigation feedback constitutes the foundation for success.

MIES procedures are guided using Fluoroscopic imaging, which continously transmits x-rays through the patient. This provides the surgeon with a real-time visualization of their catheters and guidewires inside the human body. However, fluoroscopy increases the cancer risk for clinicians and patients \cite{mahesh2001fluoroscopy,stahl2016radiation}. Clinicians operating under fluoroscopic radiation exposure must wear heavy, cumbersome protective lead vests. Also, while endovascular instruments are radiopaque, the patient's vasculature is not visible under fluoroscopy. Thus, digital subtractive angiography (DSA) is performed intermittently during the interventions. In DSA, contrast agents are injected to temporarily visualize the vasculature. Yet these  agents are nephrotoxic, posing burden on the patient's kidneys \cite{wong2016contrast, arnautovic2024radiation}, with  consequences as drastic as acute kidney failure. Due to above concerns, non-fluoroscopic navigation techniques have gained interest in the past decade.

Magnetic resonance imaging (MRI) and ultrasound are alternative intraoperative imaging modalities. MR guided surgery offers high quality imaging, 3D visualization, and functional information, and has been used for angioplasty and stent placement \cite{nijsink2022current,kilbride2022mri}. Yet it restricts surgeon's access to the patient \cite{su2022state}, and requires costly and bespoke MR compatible instruments \cite{abdelaziz2024thermally}. 
Ultrasound offers real-time imaging without contrast agents \cite{sandhya2020randomised, he2017comparison}. It was used in EVAR \cite{kopp2010first} and angioplasty \cite{wakabayashi2013ultrasound}, but suffers from imaging artifacts and a small field of view \cite{gibbs2011ultrasound}. Besides imaging, proprioceptive sensing has also gained interest, e.g. based on Fiber Bragg Gratings (FBG). Commercially, the Fiber Optic Real-Sense system (FORS, Philips Medical, Amsterdam, Netherlands) uses FBGs in its catheters for sub-millimeter shape reconstruction \cite{megens2021shape}, shown in pre-clinical \cite{bydlon20233d} and clinical trials \cite{van2021first}. Yet, FBG systems require registration and carry a high cost.

Another approach, Bioelectric Navigation (BN), uses patient-specific electric impedance signals to localize endovascular devices inside the vascular system \cite{sutton2020biologically}. BN measures a local impedance by injecting a weak electric current into the blood stream through ring electrodes on the catheter. The measured impedance is dependent on the local geometry of the blood vessel that the catheter currently traverses, and changes as the catheter moves through different parts of the vasculature, creating a time-varying impedance signal. Before the intervention, exemplary impedance signals can be simulated based on a pre-operative model of the patient's vasculature. For some interventions like EVAR, such pre-operative scans are taken routinely in preparation for the surgery. During the intervention, the impedance signal measured from the catheter is matched with the simulated reference signals to determine the most probable location of the catheter inside the vascular system. Since BN uses a sensing concept that directly corresponds to the shape of its surroundings, it holds promise for identifying the precise location of key anatomical features intra-operatively, such as the start and end points of an aneurysm. This could potentially allow surgeons to position stent grafts without any fluoroscopy or contrast agents administered.

Recent studies in BN have repurposed off-the-shelf electrophysiology (EP) catheters to show feasibility and extend the concept \cite{sutton2020biologically, ramadani2023feature, maier2023extending}. Such catheters carry
%a pullwire for tip actuation, and
a number of electrodes near the tip, for recording of intracardiac electrograms. 
These electrodes are arranged in spacings defined in ISO medical device standards \cite{ISO10555-1:2023}. However, EP catheters are not specifically designed for BN. Their electrodes are neither distributed to optimize sensitivity to vascular features, nor in a way that allows the information from different electrodes to benefit the overall tracking task. They are also not adapted to the patient's specific vascular system, or the MIES intervention at hand, in which sometimes more than one region along the catheter needs to be tracked. For example in an EVAR procedure, the surgeon needs to ensure that the start of a stent graft is placed below the renal arteries, to prevent renal artery occlusion. But they also need to ensure that the end of the graft does not occlude the internal iliac artery, meaning that two critical locations need to be monitored at the same time during deployment. Especially the latter raises interest in tracking electrodes being distributed quite far from the tip and at bespoke locations on the catheter body. 

For any electrode-carrying catheter, two main considerations have to be taken. Firstly, the design and manufacture of the catheter body. This needs to offer mechanical support for the surgeon to navigate the device, working channels for conductive wires (to transmit signal to and from electrodes), a guidewire lumen, and space for additional sensors. Secondly, the  assembly process, including establishing a connection between the 
conductive wires and the electrodes, then securing the electrodes and wires in place on the catheter body. Commercial catheter bodies are fabricated using extrusion \cite{US9089339B2, US8540662B2}, which is limited by the precise machining needed to make a die, the effects of die swelling \cite{jin2014design,tanner1995extrudate}, and significant costs incurred during device optimization, precluding use in rapid prototyping scenarios. A similar problem arises when it comes to attaching electrodes, where commercial catheters typically mold the main body around attached electrode rings \cite{US20180055562A1, US10918832B2, US11730426B2}, which requires high precision molds, and automation to replicate the assembly.

\begin{figure*}[t!]
    \centering
     \hspace*{-0.8cm} 
    \includegraphics[width=0.95\textwidth]{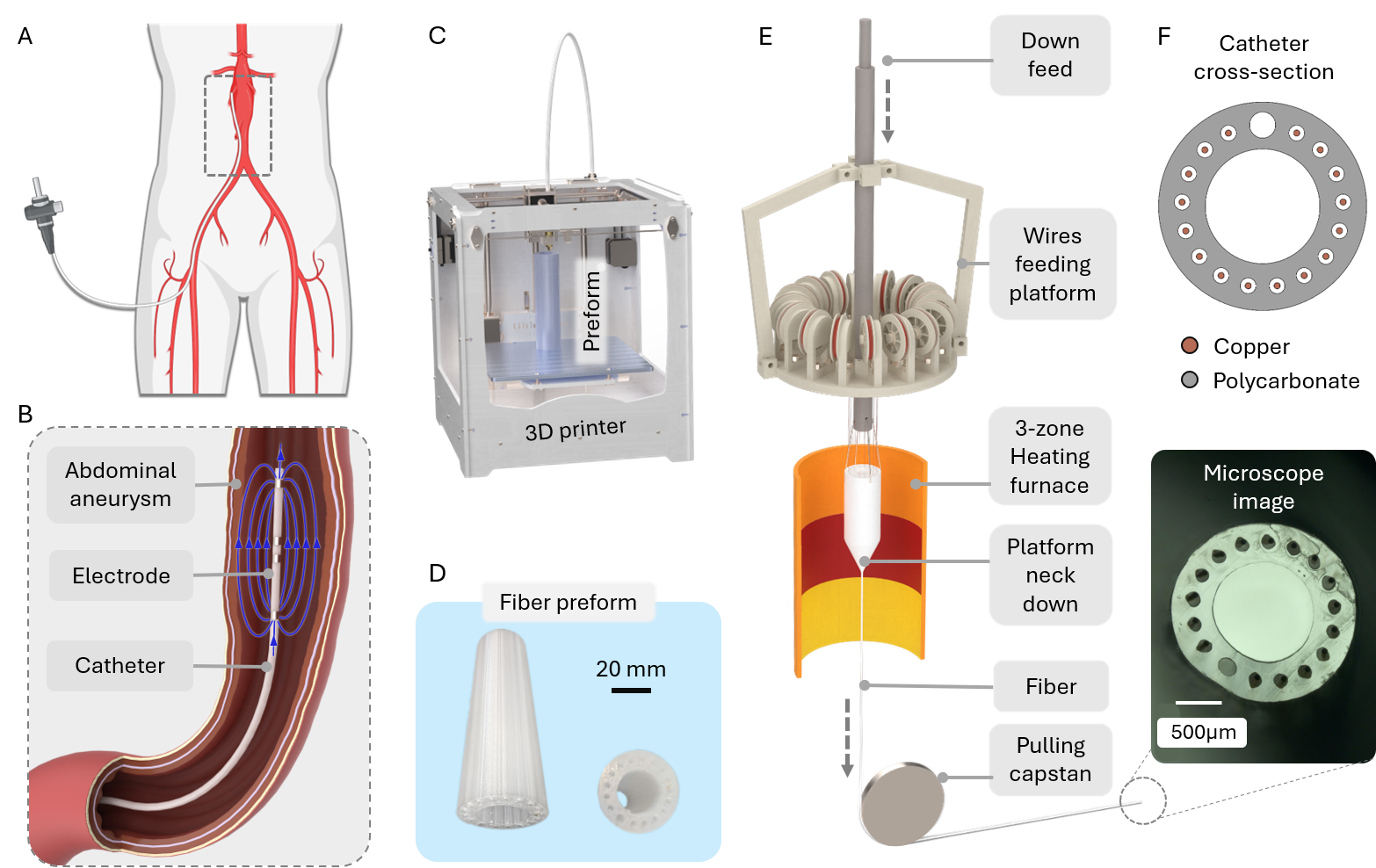}
    \caption{16-electrode catheter for Bioelectric Navigation (BN) in endovascular surgery. A) An illustration showing the device navigating inside an abdominal aortic aneurysm. B) An illustration demonstrating the underlying mechanism of Bioelectric Navigation. A weak electric current, emitted by two outer electrodes, creates an electric field sensed by the inner two electrodes resulting in a local impedance, used as a tool for localization without fluoroscopy. C) Use of 3D printing for fiber preform fabrication. D) Photograph of printed fiber preform. E) A schematic showing the thermal drawing process and the wire feeding platform designed to supply 16 copper wires in a single fiber. F) Cross-sectional view of a fabricated electrode catheter, with a central guidewire channel, 16 copper wire filled channels and 1 working channel for other sensors. }
    \label{fig:main figure}
\end{figure*}

In this work, we tackle these challenges by creating a novel pipeline for the fabrication of bespoke electrode-equipped catheters, suitable for impedance sensing (Figure \ref{fig:main figure}A,B). We begin by introducing the thermal drawing technique, capable of making highly customizable catheters in a rapid, scalable and affordable manner. We characterize the mechanical behavior of our prototypes by evaluating their tensile strength, flexural stiffness, torsional stiffness and axial stiffness, then compared these metrics with a commercial medical-grade product. The final device assembly primarily utilizes laser micromachining, needed to expose and attach copper wires to platinum electrodes. To enable this sensor for clinical use, we further introduce algorithms for real-time tracking of the catheter's location inside the aorta, for the first time fusing the concept of sensing of vascular geometric features using local impedance \cite{sutton2020biologically, ramadani2023feature} with electric motion and distance estimation \cite{maier2023extending} into a real-time catheter localization pipeline. We evaluated the system under several in-vitro conditions, from a simple phantom as proof of concept, to an anatomically realistic phantom. We then emulated the treatment of an abdominal aortic aneurysm, having an experienced endovascular surgeon navigate our device, guided in real-time solely by our proprioceptive sensing.

Our key contribution is two fold. First, expanding the thermal drawing technology to fabricate intricate, and complex catheters, enabling technologies such as BN to progress. Second, introducing a new algorithm for BN tracking, creating new paradigms for real-time guidance of catheters without fluoroscopy. 

\section{Results and Discussions}

\subsection{Design and Fabrication of Wire-Embedded Catheters}

\subsubsection{Thermal Drawing for Catheters and Fiber Design} \label{Section: Thermal drawing}

We used thermal drawing to create scalable multi-lumen catheters in a cost-effective manner. The preparation steps include 3D printing a fiber preform, which is a macroscopic version of the catheter design (Figure \ref{fig:main figure}C,D). The preform, 4~cm in diameter and 17~cm in length, is 3D printed (Ultimaker3+ Extended, Utrecht, Netherlands), allowing rapid fabrication of arbitrary cross-sectional designs (Figure \ref{fig:main figure}D). We selected polycarbonate (PC, Tg=$112^o-113^oC$, $\alpha$ = $2.1 \times 10^{-5\: o}  C^{-1}$) as the catheter material, as it is 3D printable, bio-compatible, has high mechanical strength, and dimensional stability \cite{gomez2021mechanical,abdelaziz2024fiberbots, demircali2025fabrication}. The preform was mounted onto the preform holder, a 60~cm long polyetheretherketone (PEEK) rod. Additionally, a wire-feeding platform was assembled and secured to the preform holder above the preform. This platform contains 16 wire bobbins that feed copper wires into the preform (Supporting Information Section~1.1)

During the drawing process, the preform is fed downwards into a cylindrical heating furnace of a thermal drawing tower, and heated above its glass transition temperature. The furnace has three heating zones, and temperatures of 120/190/85 $^o C$ were applied to the top, middle, and bottom zones. These temperatures were initially selected based on literature \cite{demircali2025fabrication}, then fine-tuned over several iterations for optimal diameter consistency and cross-section integrity of our design. After the preform is heat-softened into a viscoelastic state, it is continuously drawn from the bottom of the furnace using a rotating capstan (Figure \ref{fig:main figure}E). This drawing process produces thin, elongated fibers, while maintaining the preform's cross-sectional profile. The steps of the drawing process and the thermal drawing tower setup are shown in Supporting Video S1.

The resulting catheter features a multi-lumen cross-section with a 1.1mm guidewire channel in its center, and an overall diameter of 6~Fr (2~mm), which is the standard catheter size used in such procedures (e.g. \cite{us10517603b2}). Surrounding this center channel are seventeen radially distributed side channels, sixteen of which are for copper wires (160~µm, for 50~$\mu m$ copper wires) and one for multi-purpose use (250~µm), like adding an additional FBG sensor or a pullwire (Figure \ref{fig:main figure}F). We selected 16 electrode wire channels as this is the maximum that can fit into the preform while ensuring sufficient spacing to prevent coalescence during the draw. This is a compromise between spatial constraints, ease of wire handling during manufacture, wire resistance, and electrode count. Figure \ref{fig:Mechanical Characterisation}A shows a number of catheter bodies fabricated in a single draw. Changes in the fiber diameter during the thermal drawing process are presented in Figure \ref{fig:Mechanical Characterisation}B. Details regarding the parameters of the drawing process can be found in Supporting Information Section 1.2.

\subsubsection{Laser Surface Profiling}

Manipulating a catheter in the human vasculature may create high axial and bending forces, risking vessel perforation or dissection \cite{ihn2018complications, ryu2011vascular}. To mitigate this, we profiled the catheter surface between the electrodes with a femtosecond laser. The profile is made of simple straight slots (detailed in Section \ref{Section: Methods-laser}), with depth optimized to avoid damaging the copper wires, while improving the catheter's flexibility (Figure \ref{fig:Mechanical Characterisation}C). 

\subsection{Mechanical Characterization of Wire-Embedded Catheters}

\begin{figure*}
    \centering
    \includegraphics[width=0.75\textwidth]{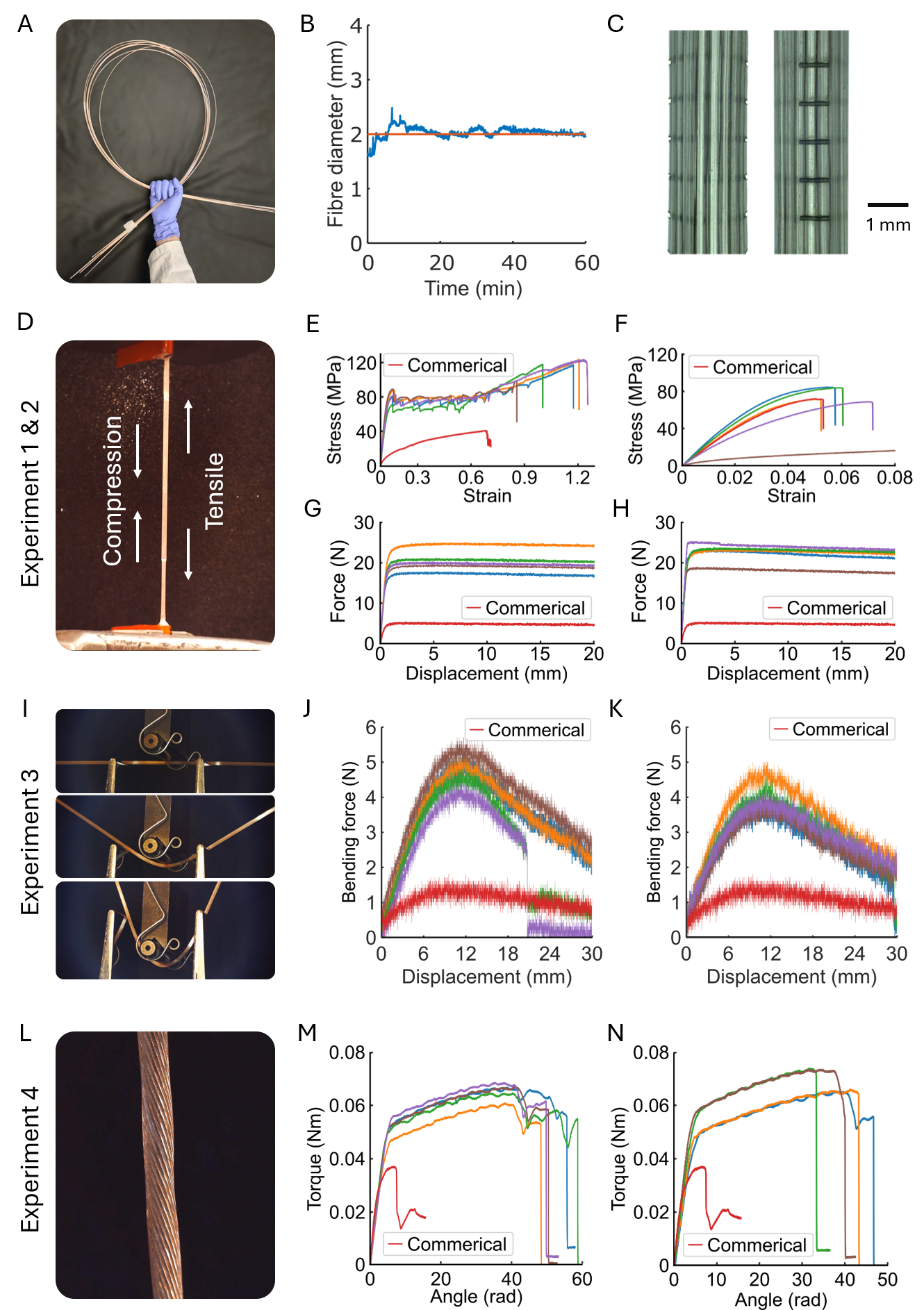}
    \caption{Mechanical characterization. A) Photograph of catheters produced in a single draw. B) Time evolution of catheter diameter during thermal drawing. C) Microscope image (x100) of the catheter's side view showing a femtosecond laser-machined surface profile. D) Illustrations of tensile and compression testing. E) and F) Force-displacement curves from compression tests on catheters without - E) and with - F) laser profiling. G) and H) Tensile test results on catheters without - G) and with - H) laser profiling, where only the first portion of the stress-strain curve (until up to 0.08 strain) for the commercial sample is displayed in H). I) Illustration of 3-point bending test. J) and K) The bending force-displacement relationships for catheters without - J) and with - K) laser profiling. L) Illustration of torsional testing. M) and N) The torque-twist angle relationships for catheters without - M) and with - N) laser profiling.}
    \label{fig:Mechanical Characterisation}
\end{figure*}

Endovascular catheters must withstand a range of mechanical stresses, including kinking, bending, and axial forces in the patient's body. In addition, thermal drawing involves non-uniform heating and deformation of a polymer preform under high tension, which may lead to substantial changes in its microstructure and residual stresses \cite{richard2022unraveling}. Thus, it is essential to assess the drawn catheter's mechanical performance.

We conducted a comprehensive comparison between a commercial 6 Fr catheter (Merit Medical Impress Selective Catheter, Utah, USA) and our prototype with and without laser profiling. We chose the following key mechanical characteristics: Young's modulus, yield stress and tensile stress (Experiment 1: Tensile test), flexural stiffness (Experiment 2: Three point bending test), torsional stiffness (Experiment 3: Torsion test), and axial stiffness (Experiment 4: Compression test). Details of catheter sample preparation are discussed in Section \ref{Section: Mech Test prep}. We tested 5 samples per trial (S1-S5), and one commercial sample (REF). Details regarding metrics computation, and the experimental setups are presented in Supporting Information Section 2, and Figure S3. The computed mechanical metrics are presented in Supporting Information Section 3, Table S1.  All 4 mechanical characterization experiments are shown in Supporting Video S2.

The Young's modulus, yield stress, and tensile stress reflect the robustness of the catheter to interaction forces during surgery. Excessive forces between catheter and anatomy may cause kinking or fracture \cite{sadler2023cardiac, malik2020knot}. We evaluated this with a tensile test (Figure \ref{fig:Mechanical Characterisation}D). The stress-strain relationship for each sample is shown in Figure \ref{fig:Mechanical Characterisation}E, F. Whether profiled or not, our catheter exhibited higher Young's modulus, tensile stress, and yield stress than the commercial one. This suggests superiority of PC in withstanding kinks, even under high forces - likely due to molecular chain entanglement \cite{zheng2023molecular, chan2019entanglement}. Unprofiled samples exhibited an increase in measured stress, elongating plastically but hardening at high strains. While the profiled prototype is brittle, due to stress concentration at the profiles, in practice, surgeons are exerting forces of 0.5-2.5N \cite{rafii2017objective}, far below the failure stresses measured. Thus, our profiled catheter satisfies this task.

Axial stiffness defines how translation of the proximal end is mapped to translation on the distal tip of the catheter. High axial stiffness transmits forces well, but raises concerns regarding vessel perforation, requiring a compromise. To investigate this, we used a compression/buckling test, where we measured the axial forces when a secured catheter is pushed from above (Figure \ref{fig:Mechanical Characterisation}D). Test results are presented in Figure \ref{fig:Mechanical Characterisation}G, H. According to clinical studies, the average force to perforate an aorta is around 2-2.5~N \cite{wylie2009therapeutic}. Through compression tests, the maximum axial force our catheter can exert is 20.72~N/22.38~N for an unprofiled versus profiled sample, which is higher than the commercial catheter (5.36~N). Results are echoed when comparing the axial stiffness. Thus, while our catheter has less chance of buckling, it is stiffer. In both cases, the surgeon must limit their handling forces when translating the device axially into the body. In the  future, atraumatic tips made of a soft elastomer may reduce the stiffness and maximum axial force.

Flexural stiffness represents the catheter's resistance to bending under perpendicular forces. A balance must be struck between a soft and flexible catheter that is necessary for safe, atraumatic navigation, and a stiff catheter to support the delivery of instruments such as stent grafts. To investigate this, a three-point bending test was carried out. 
This test involves supporting the sample on a two-point base, and applying a force at the midpoint of the catheter (Figure \ref{fig:Mechanical Characterisation}I).
Generally, removing more material from the catheter's cross section is expected to reduce the stiffness. From Figure ~\ref{fig:Mechanical Characterisation}J) and laser-profiled (Figure ~\ref{fig:Mechanical Characterisation}K, compared with the unprofiled samples, the laser profiling reduced the bending forces by 20$\%$ and the flexural stiffness by 30$\%$. The flexural stiffness of the profiled sample (0.771~kNmm$^2$) is comparable to that of commercial products (0.643~kNmm$^2$). Optimizing the profile design could reduce this further.

The resistance of a sample to twist is quantified by torsional stiffness. For surgeons, it is important how rotation applied at the proximal end translates to the distal end. We measured the torsional stiffness by fastening one end of the catheter to a servo motor, and keeping the other fixed to a force-torque sensor (Figure \ref{fig:Mechanical Characterisation}L). The sample is then twisted until it fractures or kinks. Figure \ref{fig:Mechanical Characterisation}M, N shows that the commercial catheter deforms elastically until 5 radians (similar to our prototype), then kinks at approximately 15 radians, which is around 1/4 of what is observed from the thermally drawn device. The profiled samples typically fail at a lower angle of rotation compared to the unprofiled case, suggesting that the profiling increases stress concentration and creates premature failure. 

\subsection{Electrode Assembly} \label{Section: Electrode Assembly}

Assembling electrodes onto a 16 wire catheter body involves creating small windows on the catheter surface to access copper wire channels (Figure \ref{fig:Electrode assembly}A), establishing a robust electrical connection via laser spot welding (Figure \ref{fig:Electrode assembly}B), fastening the electrode onto the catheter (Figure \ref{fig:Electrode assembly}C), and covering the catheter body with a layer of heat shrink that is subsequently selectively removed to expose the electrodes (Figure \ref{fig:Electrode assembly}D).

\begin{figure}[t!]
    \centering
    \includegraphics[width=0.93\linewidth]{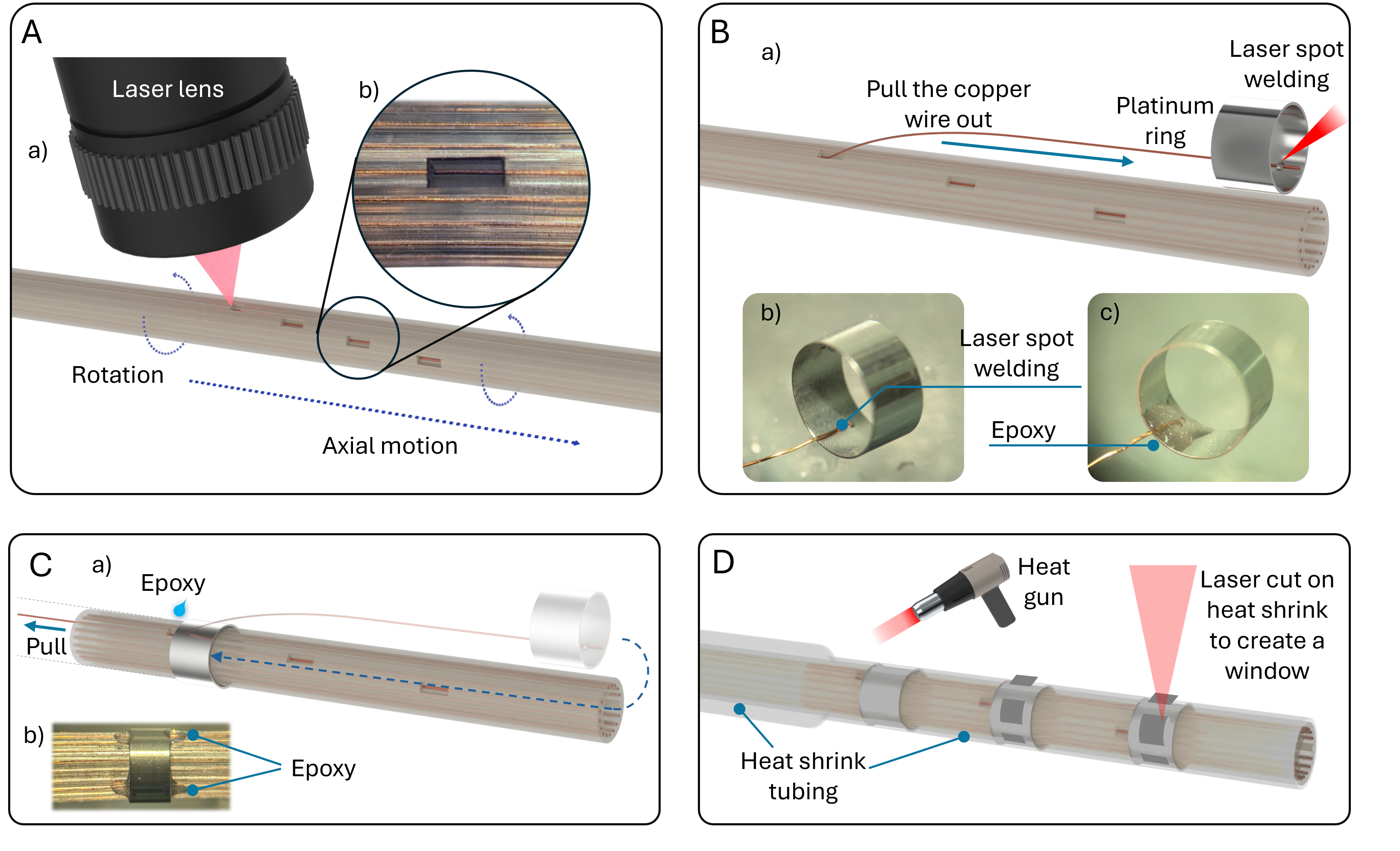}
    \caption{Detailed schematic for electrode assembly. A) Using femtosecond laser micro-machining to open windows on catheter surface. B) Laser spot-welding to fasten the electrode to a freely moving exposed wire, with epoxy to maintain robust connection. C) Assembly of the electrode on the catheter via sliding and gluing. D) Assembly of the final insulation layer using a medical grade heat shrink, with windows at the electrodes to allow for electrical conductivity.}
    \label{fig:Electrode assembly}
\end{figure}

\subsubsection{Femtosecond-Laser Micro-Machining for Wire Exposure}

The center-to-center distance of adjacent wire channels is 280~$\mu m$, with 140~$\mu m$ between channel and catheter surface. Precisely cutting windows into such a catheter that selectively expose single wires is challenging with standard CNC milling. Thus, a femtosecond laser micro-machining system was chosen, due to its high precision and reliability. It was combined with a custom-made rotary and linear stage to position the catheter under the laser lens. 16 windows of size 1 by 0.1~mm were cut along the catheter (Figure \ref{fig:Electrode assembly}A-a)), exposing the copper wires for further connection (Figure \ref{fig:Electrode assembly}A-b)). The laser system is detailed in Supporting Information Section 4.1. Laser settings are presented in Experimental Section \ref{Section: Methods-laser}.

\subsubsection{Laser Micro-Welding for Electrode Assembly}
Robust and mechanically sound contact between the electrodes and copper wires is critical. Poor contact may lead to high resistance, noisy impedance signals, or connection breaks. We applied laser micro-welding for connection, as it retains a low profile, low-resistance connection during its lifetime, while adhesives and liquid metal epoxies may peel and lose connection over time \cite{mills1964comparison, li2021laser}. 
Each copper wire was pulled out of the catheter for a few centimeters, through the corresponding laser-machined window; placed on the inner face of the ring electrode, and subjected to the welding laser until a weld visibly formed (Figure \ref{fig:Electrode assembly}B-a)). A small layer of epoxy resin was used to secure the contact (Figure \ref{fig:Electrode assembly}B-b),c)). Then, the connected electrode was threaded onto the catheter, slid into its desired location, and secured in place with additional epoxy (Figure \ref{fig:Electrode assembly}C). For additional seal and protection of the electrode's rim, a thin layer of medical-grade heat-shrink was used to cover the entire catheter. The catheter was again laser machined to remove the heat-shrink on top of the electrodes, exposing several square-shaped windows per electrode (Figure \ref{fig:Electrode assembly}D). On the catheter's proximal end, wires were exposed and soldered to a 3D-printed handle (Supporting Information Section 1.3, and Figure S2). 

\subsubsection{Electrode Configuration in Catheter Design}
We spaced the electrodes on our catheter based on clinical needs and the principles of BN. The goal was to create multiple sets of electrodes that can each serve as local vascular shape sensors. As suggested in \cite{ramadani2023feature,maier2023extending}, we used a tetrapolar configuration (Figure \ref{fig:electrode_configuration}A) as a base sensing unit to detect vascular geometric features, like changes in vessel diameter. Here, an alternating current (AC) is generated between a pair of two outer electrodes (the injection electrodes, pink in Figure \ref{fig:electrode_configuration}A), and the corresponding electric voltage is measured between a pair of inner electrodes (the sensing electrodes, green in Figure \ref{fig:electrode_configuration}A). Given known current and measured voltage, a local electric impedance can be calculated. 

When combined with an additional electrode at a stationary location inside the bloodstream (e.g. at the vascular access sheath), the three distal electrodes of a tetrapolar configuration can also be used to estimate the distance and direction (forward or backward) the catheter travels into the blood vessel. This is referred to as a displacement sensor \cite{maier2023extending}. A tetrapolar configuration can be used both as a displacement sensor and a vascular shape sensor at the same time.  One more electrode shortly behind the distal injection electrode helps to mitigate displacement estimation errors \cite{maier2023extending}, leading to a 5-electrode setup (Figure \ref{fig:electrode_configuration}B).

We added more electrodes to create additional sensor configurations (justification of electrode distributions see Supporting Information Section 4.2). This can later allow for sensor fusion by combining tracking estimates from multiple sensing units.

\begin{figure}[t!]
    \centering
   
    \includegraphics[width=0.95\linewidth]{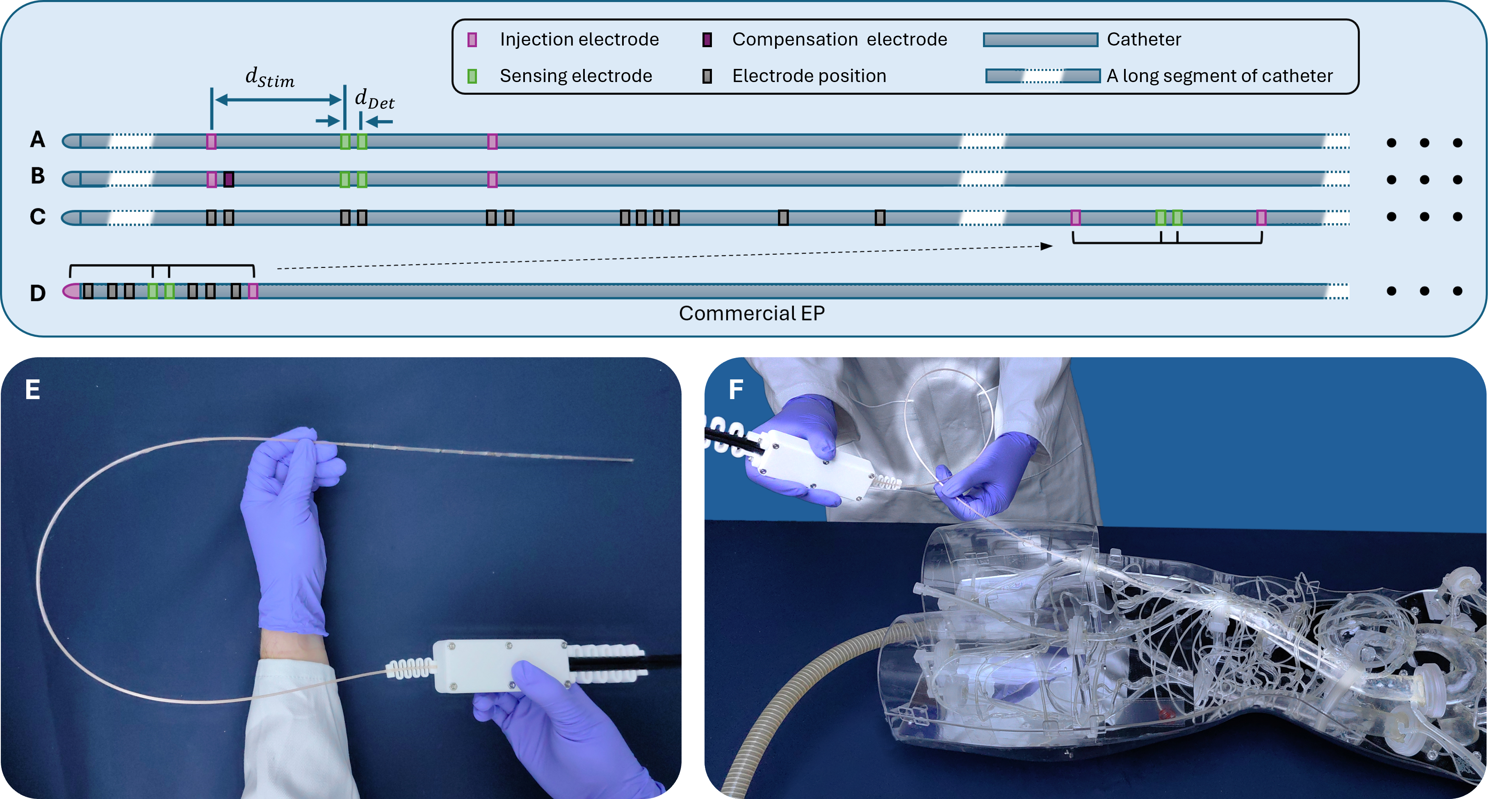}
    \caption{Schematic of electrode distribution. Sensing electrodes are indicated in green, injection electrodes in purple. The electrodes outlined in a darker shade of purple are compensation electrodes for the displacement estimation. A-C represent our prototype, while D represents a configuration used on the commercial EP catheter. 
    }
    \label{fig:electrode_configuration}
\end{figure}
In some endovascular procedures, the surgeon has to know the catheter’s relative position with respect to multiple vascular landmarks. In EVAR, the surgeon needs to ensure that the aortic end of the stent graft does not occlude the renal artery ostia. Yet, they also need to ensure that the other end of the graft has moved past the bifurcation of the common iliac artery, to avoid occlusion of the internal iliac artery.  To aid in such a case, we added one tetrapolar configuration more proximally, 195 mm behind the first distal injection electrode (this is a patient-specific parameter; rationale see Supporting Information Section 4.3). In our AAA phantom, when the surgeon places the distal catheter electrodes at the renal ostium, the proximal configuration allows the surgeon to confirm that a defined portion of the catheter has passed the iliac bifurcation.
This helps to confirm a catheter location where both ends of the stent graft are safe to be deployed without occluding any branching arteries. With this configuration, our final electrode distribution is as shown in Figure \ref{fig:electrode_configuration}C.

A tetrapolar configuration is defined by the inter-electrode distances $d_{stim}$ and $d_{det}$ (Figure \ref{fig:electrode_configuration}A).
Based on works on impedance-based estimation of vessel diameter \cite{kassab2004measurement, kassab2005novel, kassab2004measurement, woodard1989effect}, we chose 29~mm as $d_{stim}$, and 3~mm as $d_{det}$ (details in Supporting Information Section 4.4). Both are chosen based on our vascular phantom geometry, i.e. patient- and intervention-specific vessel diameters. 
To compare our catheter to a commercial EP catheter (Boston Scientific EP XT 6~Fr decapolar), which has smaller electrode distances, we made $d_{stim}$ of the most proximal configuration (the \textit{reference configuration}; Figure \ref{fig:electrode_configuration}C) identical to a tetrapolar configuration of the EP's 1\textsuperscript{st}, 5\textsuperscript{th}, 6\textsuperscript{th} and 10\textsuperscript{th} electrodes (Figure \ref{fig:electrode_configuration}D), with $d_{stim}$ as  18~mm. The final prototype, along with the handle, is shown in Figure \ref{fig:electrode_configuration}E and 4F.

\subsection{Electrical Characterization of Assembled Electrode Catheter}

\begin{figure}[t!]
    \centering
    \includegraphics[width=0.95\linewidth]{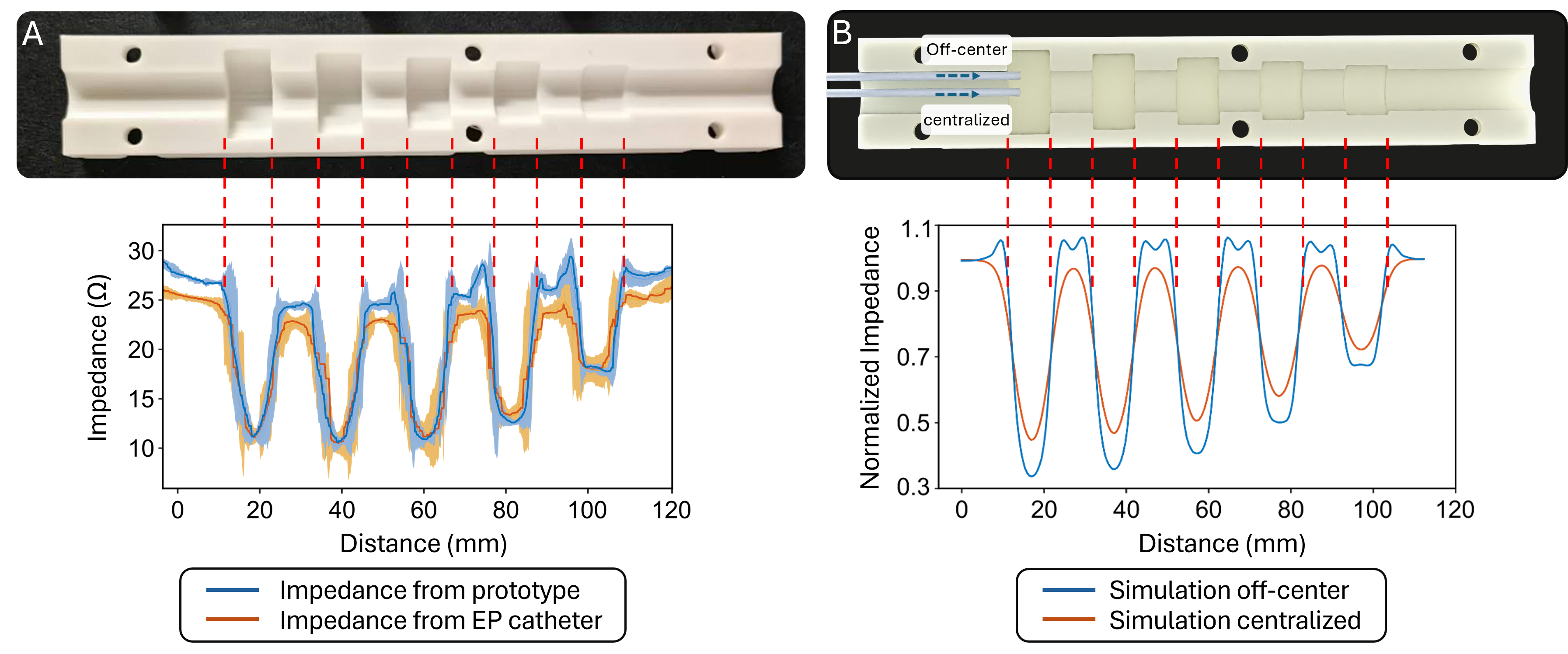}
    \caption{In vitro and in silico evaluation of impedance sensing in a simple phantom. (A) Comparison of recordings from our prototype and commercial EP, in a simple 3D printed phantom. Shown are the phantom's cross-section, and the recordings depicted as mean and standard deviations (visualized as 3 standard-deviation interval; i.e. 99.7\% interval) of 5 pullbacks (B) FEM-simulation results. Shown are 3D FEM model (phantom lumen) and two simulated signals comparing catheter on centerline with catheter off-centerline (close to wall). Catheter positions are indicated by two catheter tips on the left.}
    \label{fig:resistance_and_simple_phantom}
\end{figure}

\subsubsection{In-Vitro Evaluation with Simple Geometry Phantom}
Each electrode was verified to be connected to the handle with low resistance (Supporting Information Section 5, Figure S5).
To assess the performance of our prototype for recording vascular features for BN, we 3D-printed a simple phantom consisting of a straight tube, with multiple bulges of increasing diameter (cross-section in Figure \ref{fig:resistance_and_simple_phantom}A).  We recorded tetrapolar impedance signals using both our prototype and the commercial EP catheter, using our prototype's \textit{reference configuration} for comparability. 
For each catheter, we mounted an electromagnetic (EM) tracker next to the sensing electrodes. Impedance signals from five pull-backs were recorded, at the sampling frequency of the impedance analyzer (Experimental Section 4.9). EM coordinates were used to resample all signals at intervals of 0.25~mm, making them independent of the speed of pullback. Average signals and standard deviations from each catheter are shown in Figure \ref{fig:resistance_and_simple_phantom}A.

Signals from our prototype and the commercial EP were highly comparable. Local impedance measured with a tetrapolar setup is expected to be inversely proportional to a vessel's cross-sectional area \cite{svendsen2013accurate}. 
Signals from both catheters showed this inverse relationship. Compared to the EP, our prototype measured larger differences between the aneurysmal and normal sections, a slightly higher average resistance, and sharper changes with small peaks at the start and end of bulging sections. We investigated this in a finite element method (FEM) simulation (experimental section 4.10). We simulated the signal from an electrode catheter on two trajectories: along the centerline of the phantom, and parallel to the centerline but close to the wall. 
Simulations showed that peaks originate from a catheter being off-center (close to the wall) while moving past the bulging sections (Figure \ref{fig:resistance_and_simple_phantom}B). Off-center positioning also caused larger signal differences between bulging and normal sections. In practice, we observed that our prototype, due to higher stiffness than the EP, was pressed closer to the wall due to catheter-phantom interaction.
We thus attribute the difference in signals to off-center alignment of our prototype. Smaller additional influences could stem from electrode spacing deviating from the intended spacing due to manual placement during assembly. Overall signal shape and features like the phantom diameter changes, were similar in signals from both catheters, showing that our prototype is suitable for recording BN features and comparable to a commercial device. 

\subsubsection{In-Vitro Evaluation with Anatomically Realistic Phantom}\label{subsubsec:in_vitro_abdominal_phantom}
We evaluated the prototype in a commercial, silicon model of an abdominal aortic aneurysm (Elastrat Sàrl. Figure \ref{fig:AAA_and_distance}A; preparation see Experimental Section 4.13). 
\begin{figure}[h!]
    \centering
   \includegraphics[width=0.95\linewidth]{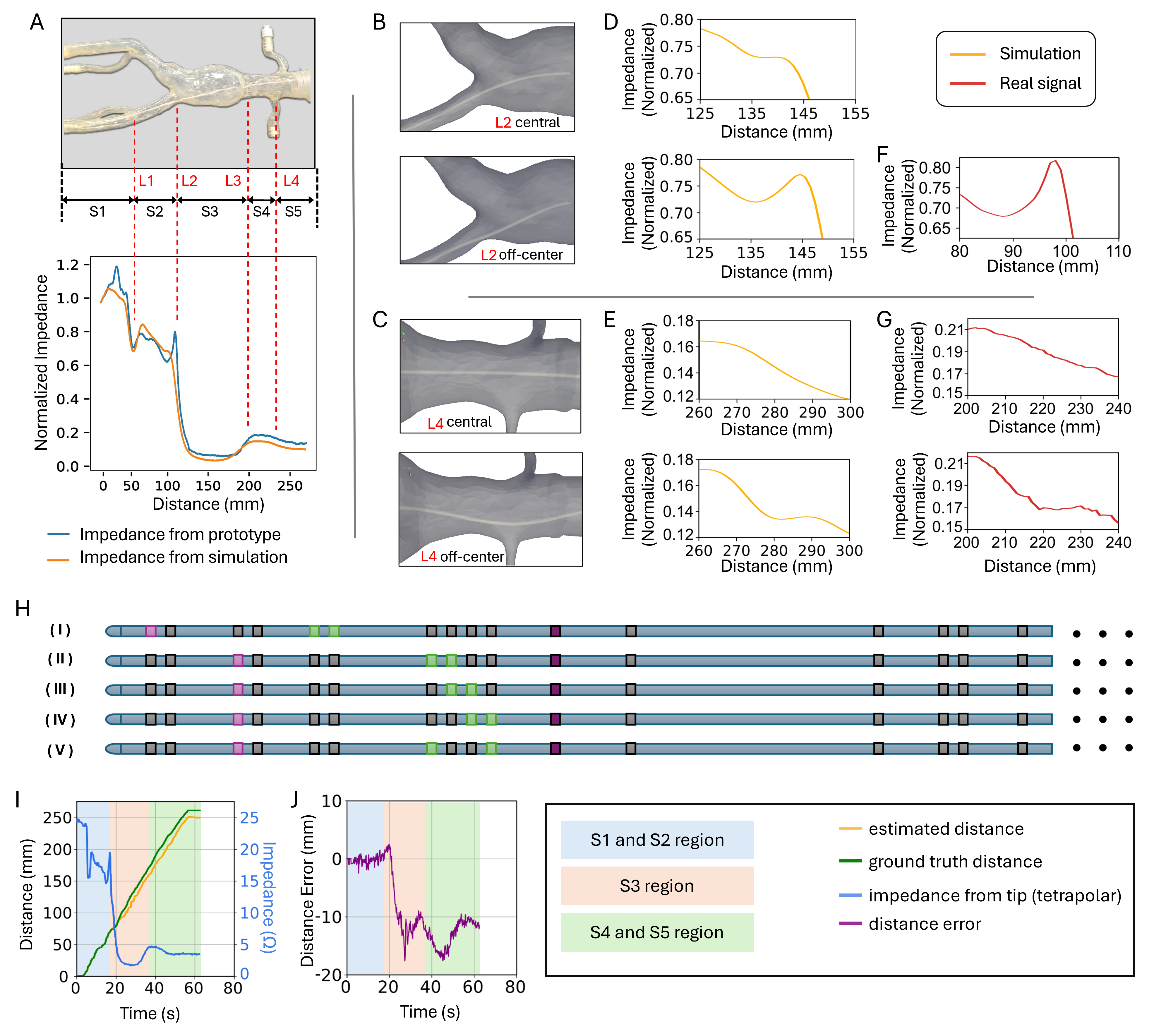}
    \caption{In-Vitro and In-Silico characterization of impedance sensing for our prototype in an anatomically realistic phantom (A) Tetrapolar recordings in anatomically realistic silicon phantom of abdominal aortic aneurysm. Shown are the phantom with sections comprising the iliac artery (S1, S2), aneurysm (S3), healthy aorta (S4, S5), and multiple locations of interest (L1 through L4). Also shown is an exemplary signal recorded in the phantom by our prototype, compared with a signal from an FEM simulation based on the centerline of the phantom. (B,C,D,E,F,G) Radially dependent features are found at (B) the start of the aneurysm and (C) at the bifurcation of the renal arteries. Shown are (B,C) the FEM geometry with the catheter inside the phantom for both locations, each time with a catheter on the centerline and a catheter off-centerline, (D,E) the simulated signals corresponding to the FEM geometries, and (F,G) a comparable signal obtained from our prototype. Top rows show centralized positions, bottom rows show off-center catheter positions. (H) different electrode configurations used for distance estimation. Purple: injection electrodes (light purple: used for distance tracking; dark purple: used to complete tetrapolar tip configuration). Green: sensing electrodes (I, J) Exemplary distance estimation results, showing ground truth and estimated distance, impedance signal obtained from the given tetrapolar pair, and error in distance as functions of time.}
    \label{fig:AAA_and_distance}
\end{figure}
We compared recordings from our prototype with a simulated signal from a centerline-based FEM simulation (Experimental Section 4.10). The recording closely followed the simulated signal (Figure \ref{fig:AAA_and_distance}A). A live recording is shown in Supporting Video S3. Both recording and simulation displayed the expected decreases in magnitude with increasing vessel cross-sectional area (most visible in the aneurysm, section S3 in Figure \ref{fig:AAA_and_distance}A), and exhibited a drop when the catheter passed the bifurcation of internal and external iliac arteries (location L1 in Figure \ref{fig:AAA_and_distance}A). 
Notably, at the start of the aneurysm (L2), our prototype recorded a peak not predicted by simulations. Also, the bifurcations of the renal arteries (e.g. L4) were neither detectable in the simulated signal, nor in our recordings. We found both effects to depend on the catheter's radial position in the vessel.
The peak at L2 stems from the catheter being close to the wall of the phantom while passing this location. We compared simulations of a centralized catheter and a catheter near the wall, and could reproduce the peak (Figure \ref{fig:AAA_and_distance}B, D). The renal bifurcation was also visible when simulating an off-centerline catheter in this location, and we could replicate this in real recordings by forcing the catheter towards the wall (Figure \ref{fig:AAA_and_distance}C,E,G).

Recently, Maier et al. proposed to estimate the distance a catheter traveled into a blood vessel based on an electric field between electrodes on the catheter and an electrode in a stationary location \cite{maier2023extending}. Their approach produces per-timestep displacement and direction estimates, which can be summed up to approximate the total catheter motion relative to the catheter’s location at the start of the measurement.
We evaluated our prototype on this task. As stationary electrode, we used an electrode of an EP fixed to the start of the phantom (start of S1 in Figure \ref{fig:AAA_and_distance}A). Our prototype has multiple electrode configurations suitable for distance tracking, and thus, we evaluated distance estimation accuracy on a subset of them (Figure \ref{fig:AAA_and_distance}H). We mounted an EM tracker next to the sensing electrodes to obtain a ground truth catheter location and registered the EM coordinate system to our phantom model. The ground truth distance was calculated by projecting EM tracking coordinates to the vessel centerline and calculating the distance between start and end locations along the centerline. The distance error is calculated as the difference between ground truth and the electrically estimated distance.

For each configuration, we moved the catheter through the whole phantom and recorded signals from six forward traversals and six pull-backs. Full results of the distance estimation, and a baseline  \cite{maier2023extending}, are shown in Supporting Information Section 6.1, Table S2. Mean distance estimation errors ranged between +1.1\% and -10.2\% of the traveled distance, with all except one configuration being off by 6.6\% or less. For most configurations, errors were comparable or slightly larger than those of Maier et al. (-3.4\% relative error). Configuration (II) was the most accurate at 1.5 \% for forward and 1.1\% for backward motion. Except for configuration (II), all configurations underestimated the traveled distance (negative mean error), similar to  \cite{maier2023extending}. Standard deviations were higher than reported by Maier et al., up to a factor of 4.4. All configurations exhibited differences between forward and backward recordings, up to 6.9\% for configuration (IV).

Differences between configurations, and to Maier et al., can be attributed to multiple factors. Variations between intended electrode spacing and actual spacing might exist. With an intended $d_{det}$ of 3.0~mm, a 0.1~mm error in spacing would cause a 3\% mean error based on the estimation concept \cite{maier2023extending}. Also, our phantom is more complex than Maier et al.'s. It features higher variations in diameter and sections with much larger vessel diameter, which permits catheter motion far from the centerline.
Motion components perpendicular to the centerline are not modeled in the formula for distance estimation, so this could cause further errors. This could both explain the higher standard deviations in our results, as well as the differences between forward and backward recordings: Due to contact with the vessel wall, the catheter assumes a different path through the phantom in forward runs, compared to pullbacks. 

We also investigated the evolution of the error within different parts of the phantom. Generally, estimation errors varied across phantom sections, with distance being slightly overestimated inside S1 and S2, but stronger underestimation inside the larger-diameter sections S3, S4, and S5 (strongest at the start of S3). One example is shown in Figure \ref{fig:AAA_and_distance}I,J (more cases in Supporting Information Section 6.2, Figure S6).

\subsubsection{Real-Time Navigation Experiment}
Finally, we conducted a real-time navigation experiment. Prior work used the DTW algorithm to localize a catheter through BN \cite{sutton2020biologically}. Yet, they only a) classified in which branch of the vascular tree the catheter was without estimating the precise location inside a branch, b) only did so post-hoc after a full traversal of the vascular tree, c) did not allow back- and forward motion of the catheter, and d) never had a user manipulate the catheter based on real-time feedback from BN. We address these limitations, providing a) precise, continuous localization of the catheter in the vascular tree, b) in real-time, during catheter motion, c) allowing to pull-back and re-advance and d) evaluated our real-time navigation with an expert surgeon.

We use a tetrapolar configuration on the distal part of the catheter to obtain, per timestep, both an impedance value and a distance estimate. These are fused into a position estimate, in a pipeline depicted in Figure \ref{fig:tracking_and_surgeon_test}. Two users - an endovascular surgeon, and a lay person - then navigated the catheter to a target location inside the phantom based only on the position estimate from our system.

\begin{figure}[t!]
    \centering
    \includegraphics[width=\linewidth]{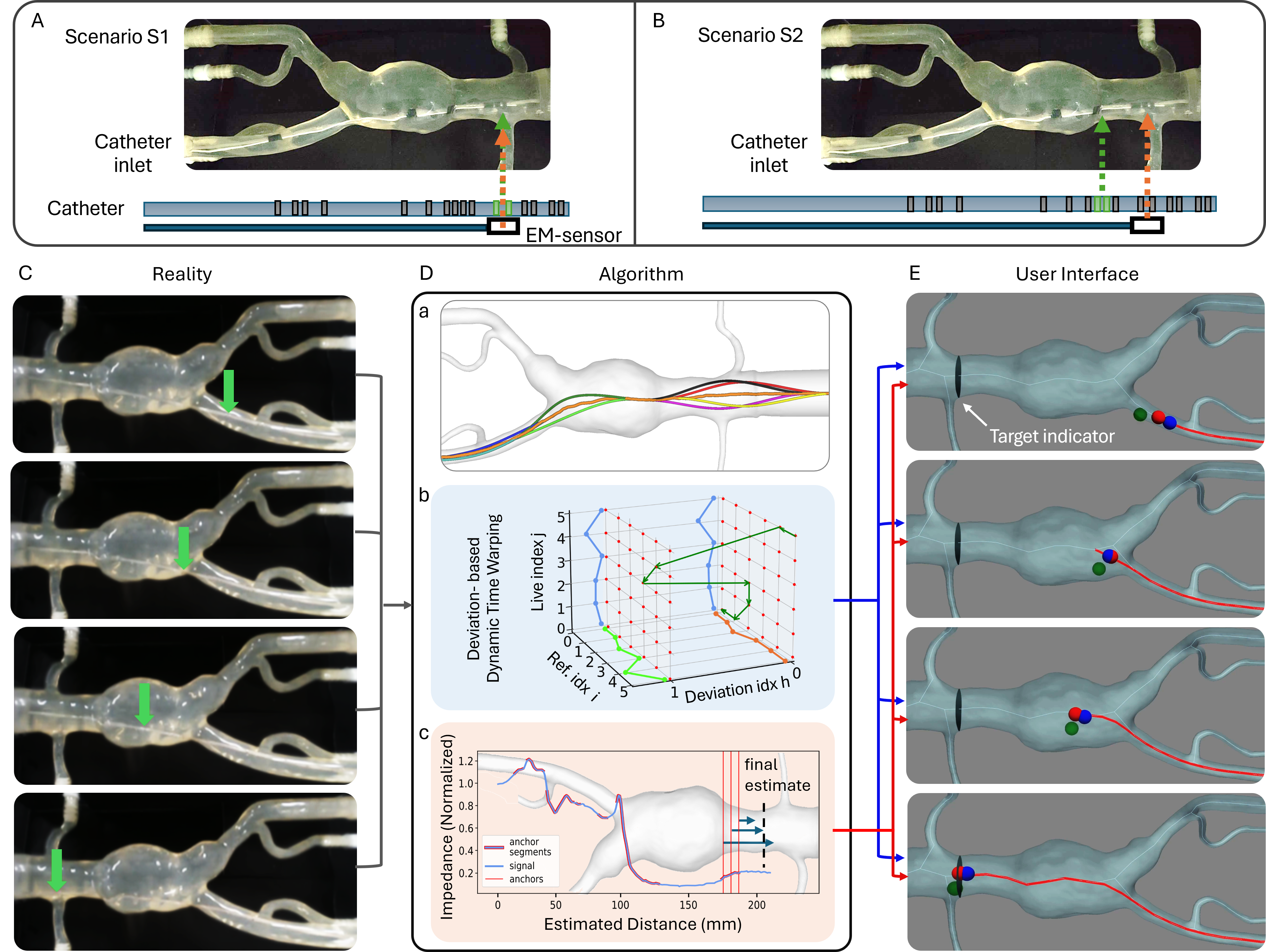}
    \caption{Live tracking setup and pipeline. (A,B) the catheter configurations and target locations of scenario S1 and S2. (C) Exemplary set of catheter locations visited during the experiment. (D-a) Catheter trajectories used for FEM simulation. (D-b) Illustration of devDTW to match one live signal to multiple deviating reference signals. (D-c) Illustration of approach A2 showing anchor region, anchor selection and final location estimation. (E) User interface showing the estimated catheter locations as spheres in the 3D phantom model, and an intersecting plane to indicate target location. Note that (E) shows two spheres (blue, red) for illustration of the two different approaches; in reality, at any time only one of the spheres was shown to the user. The green sphere shows the ground truth catheter location, which was also hidden from the user.}
    \label{fig:tracking_and_surgeon_test}
\end{figure}

To estimate the real-time position, we compiled a database of simulated reference signals.
First, we simulated a base reference signal from a centerline of the phantom (Figure \ref{fig:tracking_and_surgeon_test}D-a, orange line). We then created additional, off-centerline trajectories, by bending parts of the centerline toward one of the bifurcations of the phantom (Figure \ref{fig:tracking_and_surgeon_test}D-a; Supporting Information Section 7). 
From these \textit{deviating} trajectories, we simulated reference signals that contain radially dependent features as discussed earlier (Supporting Information Figure S10). As each deviating trajectory differs from the centerline only in one small region, their simulated signals differ from the centerline simulation only in a corresponding small region defined by a start and end index. We stored the simulated signals, as well as start and end indices, into the database.

We then designed an algorithm to match real-time catheter signals to our database.
In general, the live catheter signal should be matched to the centerline-based simulation. Yet in regions with radially dependent features, the algorithm should be able to select between centerline-based simulation and off-centerline simulations. This accounts for the presence or absence of these features, which depends on the catheter's radial position inside the vessel.

To achieve this, we modified the original DTW algorithm into a 3-dimensional DTW, which we term deviation-based DTW (devDTW). 
Traditional DTW  matches two signals by calculating an optimal, cost-minimizing alignment between them \cite{sakoe1978dynamic}. Each value of the reference signal, with index i (1\textsuperscript{st} dimension), is matched to a value of the live signal, with index j (2\textsuperscript{nd} dimension). Any match is thus defined by a tuple (i,j), and the overall matching is a sequence of such tuples. 
DevDTW generalizes this two-dimensional matching to enable matching between a live signal and a set of N reference signals that deviate in specific regions from one ``base'' reference signal (in our case: the centerline-based simulation). Each match is a tuple (h,i,j) with an additional index $h \in \{ 0, 1, \dots, N - 1 \}$, indicating which of the N reference signals the match (i,j) refers to. The algorithm is constrained so that it can match the live signal to signals other than the base reference signal only in specific regions.
In our case, each off-centerline simulation differs from the base reference signal in exactly one region, so we enforce matching to the base reference in all other regions. This avoids matching computations where all reference signals are identical.
DevDTW is illustrated in Figure \ref{fig:tracking_and_surgeon_test}D-b and discussed in detail in Supporting Information Section 8.
Both  DTW and devDTW calculate alignments based on a local, point-wise distance metric (Supporting Information Section 8, Figure S7, S8, S9, S10). We chose a weighted sum of the absolute difference in signal value and the absolute difference in signal derivatives, which takes both signal level and shape into account \cite{gorecki2013using}. The choice of weighting factor is detailed in experimental section \ref{Section: Weighting Factor}. Before devDTW matching, the signals are normalized and re-sampled based on the distance estimates (Experimental Section 4.15).

The catheter location is determined by open ended matching. An open-ended match allows DTW to end the matching early, matching the live signal to only a leading part of the reference signal and discarding the rest.
This allows to match incomplete live signals to complete reference signals \cite{tormene2009matching}, as is the case when the catheter has only partially traversed the vessel tree. The catheter position is estimated as the last matched reference index of an open-ended match between the normalized catheter live signal and the database, using devDTW. We used both open-begin and open-end matching to compensate for the fact that the catheter's exact starting position inside the vessel is not known, and that during most of the intervention, the catheter has only partially traversed the vessel.

We compare two tracking approaches, \textbf{A1} and \textbf{A2}. Approach \textbf{A1} directly uses the last matched reference index from devDTW as the catheter location shown to the user (Figure \ref{fig:tracking_and_surgeon_test}E).
Approach \textbf{A2} adds robust feature selection. In parts of the phantom, the signal recorded during catheter traversal is nearly constant, e.g. effectively plateauing in the aneurysm. There, an open-ended match can be imprecise as small changes in signal level, e.g. normalization or simulation inaccuracies, can greatly affect the matching. A2 therefore uses regions with  high amounts of change in the catheter's live signal, termed anchor regions. These regions are expected to be matched more robustly, due to their high amount of variation. We select anchor regions based on a variance criterion (Experimental Section 4.17). The three most recent indices of the live signal that lie inside an anchor region and at least 5~mm from each other are selected as anchors (Figure \ref{fig:tracking_and_surgeon_test}D-c). We estimate the catheter position based on the location of an anchor, and the estimated distance the catheter moved since it passed that anchor, $dS$. Each anchor is a tuple of the matched indices (h,i,j). A catheter position is estimated by adding $dS$ to the reference location $i$ that the anchor was matched to. The final catheter location is the mean of the locations estimated by the three anchors (Figure \ref{fig:tracking_and_surgeon_test}D-c).

To evaluate the real-time tracking, we recruited an endovascular surgeon experienced in EVAR procedures, and one lay person. Their task was to navigate the catheter to a target location inside the phantom. During EVAR interventions, surgeons aim to place the distal end of the stent graft just below the left renal artery. To replicate this scenario, we mounted an EM tracker onto the catheter; this tracker represented the end of a stent graft. The user's goal was to place the EM tracker just below the left renal artery based on feedback from our tracking algorithms.
We tried to achieve this goal in two different scenarios. In scenario S1, the real-time catheter location was calculated based on the sensing electrodes of configuration I (Figure \ref{fig:AAA_and_distance}H) and the EM tracker was mounted next to the sensing electrodes
(Figure \ref{fig:tracking_and_surgeon_test}A). 
We occluded the user's vision on the phantom, so they could only manipulate the catheter based on our position estimates, visualized on a user interface (second half of supporting Video S3). We visualized the estimated catheter location as a sphere in the 3D phantom model, and indicated the target location as a plane intersecting the phantom's lumen (Figure \ref{fig:tracking_and_surgeon_test}E). We also highlighted the centerline between the estimated catheter tip location and the start of the iliac artery, to provide the surgeon with a proxy visualization of the catheter body, despite the fact that BN only estimates the tip location and cannot reconstruct the catheter's shape.  

In scenario S2, we kept the EM tracker at the same position (Figure \ref{fig:tracking_and_surgeon_test}B), but changed the tracked electrodes to the sensing pair of configuration III (Figure \ref{fig:AAA_and_distance}H), creating a 29~mm distance between EM tracker and tracked electrodes. The goal was still to place the EM tracker, representing the end of the stent graft, below the left renal. But to achieve this, the user needed to place the tracked electrodes at the end of the aneurysm (which is located 29~mm before the left renal target location). The rationale was that it might benefit the target accuracy to aim for an auxiliary target, with stronger BN features compared to the renal artery, even if it meant to only indirectly target the true left renal target.
The target positions and tracker locations of both scenarios S1 and S2 are shown in Figure \ref{fig:tracking_and_surgeon_test}A and Figure \ref{fig:tracking_and_surgeon_test}B, respectively.

We present results of the real-time navigation in Figure \ref{fig:live_tracking_results} (additional statistical data provided in Supporting Information Section 9, Table S3). As a small number of runs had high target errors that could be considered outliers (mostly with approach A1), we present both statistics over all runs, as well as over the best four runs to reduce outlier influence. We also used A1 and A2 to retrospectively analyze tracking accuracy for recordings from the distance estimation (Section \ref{subsubsec:in_vitro_abdominal_phantom}), with results also in Figure \ref{fig:live_tracking_results}. As BN estimates the catheter location as a point on the vessel's centerline, while an EM sensor provides 3D coordinates that may be off-centerline, tracking errors are calculated along the centerline - between BN estimate and EM ground-truth, after projecting the EM measurement onto the centerline.

\begin{figure}[t!]
    \centering
    \includegraphics[width=0.97\linewidth]{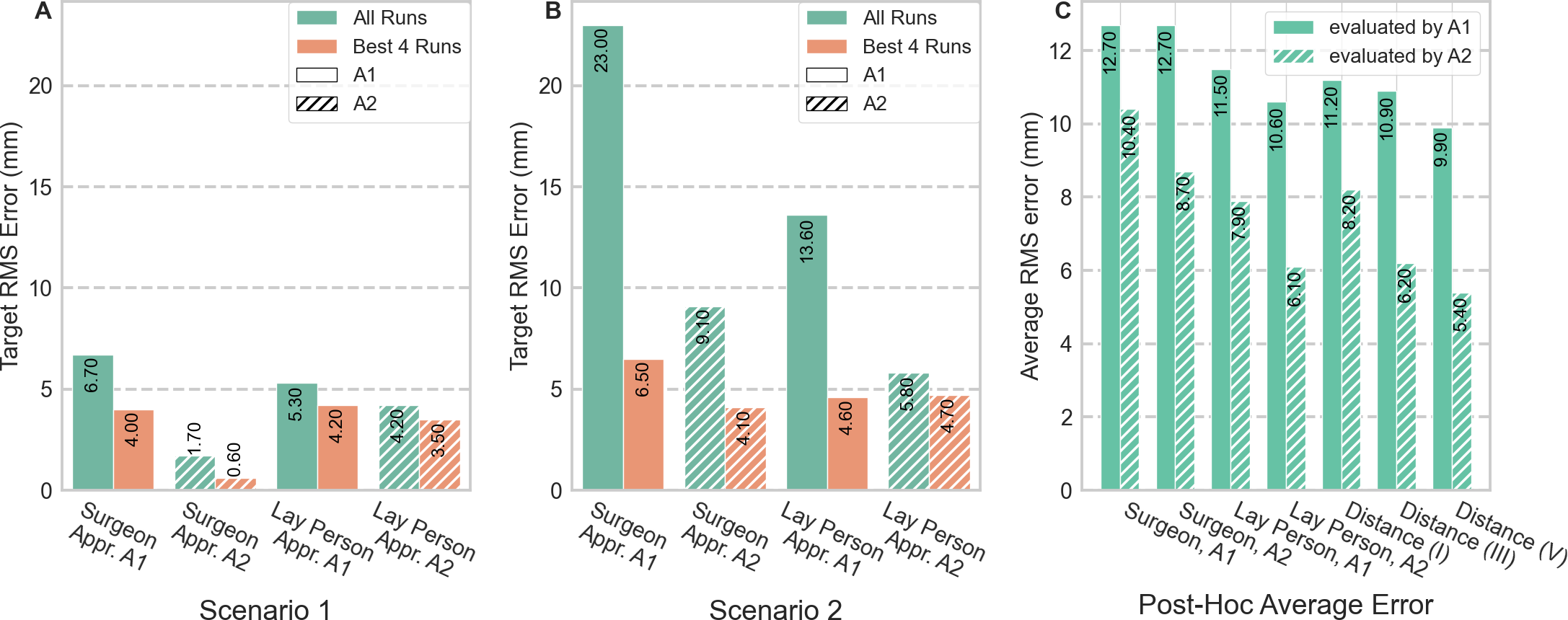}
    \caption{Tracking results. (A,B) Target RMS errors (final EM location vs. intended target) achieved by the users, either calculated across all 5 runs or excluding the worst run (best 4). (C) Average RMS error, calculated from RMS errors (EM location vs. estimated location) computed every 10~mm of ground truth motion. This can only be calculated for runs where the EM tracker was next to the sensing electrodes (i.e. Scenario S1). We also report post-hoc tracking accuracy of A1 and A2 on recordings from the distance estimation task (configuration in brackets). As this average RMS error can be calculated post-hoc, we can report results for both A1 and A2, even though users were guided either by A1 or by A2 at any given time.
}
    \label{fig:live_tracking_results}
\end{figure}

Target errors of approach A2 were consistently better than those of A1, except for scenario S2 with the lay user, where after excluding the worst run of each algorithm, both approaches worked equally well.
Indirectly targeting the renal artery (Scenario S2) was less accurate than targeting the renal artery directly (S1). An average tracking error inside the phantom was calculated from localization errors taken every 10 mm of ground truth motion. Post-hoc analysis showed that the average tracking error of approach A2 was consistently lower than that of A1. Analysis of average tracking errors on the distance recordings (rows D(I) to D(V)) shows the same trend of A2 reducing tracking errors compared to A1.
We also asked the surgeon to fill out a NASA Task Load Index (NASA-TLX) sheet, where they indicated higher-than-average mental demand (14/20), lower-than-average physical demand (7/20), low temporal demand (3/20), good performance (5/20), and very low frustration (1/20) for both approaches. A1 and A2 differed in one category: the surgeon reported that they needed rather low effort (6/20) to accomplish the task with approach A2, but more effort for accomplishing it with approach A1 (12/20). Apart from the questionnaire, the surgeon also noted that the additional, proximal tetrapolar configuration was a helpful feature. They agreed that an internal way to confirm that relevant sections of the catheter have passed the iliac bifurcation could help to verify the position of a stent graft delivery system before deployment.

\section{Discussion and Conclusion}
In this work, we presented a customizable method to manufacture bespoke electrode catheters using fiber drawing, a wire feeding platform to feed copper wires during the draw, femtosecond laser micromachining and laser spotwelding. Our prototype contained 16 copper wire channels and electrodes, a guidewire channel for over-the-wire delivery and an additional sensor channel. The single draw that produced this prototype created 21 meters of catheter body from one pre-form of 17~cm height and 4~cm diameter. This shows the capability of the fiber-drawing process for rapid prototyping. The drawing process automatically embeds wires into the catheter body. In the future, automation of laser micromachining and spot welding would be desirable, as specifically the latter needs substantial manual work and dexterity. 
The drawn catheter bodies exhibited acceptable mechanical performance, that can be further refined in future designs via laser surface profiling, e.g. to reduce axial stiffness and thus risk of perforation. Our prototype was used in over 500 insertions and pullbacks until it snapped toward the end of our study, showing its mechanical robustness. Impedance signals recorded by our prototype showed good agreement with a commercial EP catheter and FEM simulations, and catheter motion estimation results were consistent with prior work.

We also introduced a catheter tracking pipeline, for the first time fusing distance estimation with vascular features from tetrapolar impedance measurements using the BN principle. The proposed devDTW, with a refinement step based on robust feature selection, showed a tracking accuracy of 8.3~mm RMS in a real-time navigation experiment with an endovascular surgeon and a lay person. When asked to navigate the catheter to the start of the renal artery with the proposed tracking pipeline, they achieved an average target accuracy of 3~mm RMS in the runs performed with the anchor-based technique. As the first implementation of real-time catheter navigation based solely on BN, this shows the promise of the technology.

We could detect the bifurcation of the common iliac artery (location L1) in the recordings of the AAA phantom. Yet, the detectability of the renal arteries (L4), which are much smaller in diameter compared to the aorta they are branching off, was dependent on the radial position of the catheter inside the aorta. Thus, the renal arteries can be used as a localization feature if the catheter aligns well with them, but this feature is not guaranteed to be available.
Arguably, the branching of the renal artery is the most useful feature for an EVAR, as it represents the exact upper limit of the permitted stent graft advancement. Detectability of this feature might be improved through constructions that carry electrodes more radially distributed, e.g. in basket-like configurations, yet this is beyond the scope of the current study.

In future, influence of actual human tissue and of anatomy surrounding the vessels should be investigated. Surrounding tissue provides a path for electric current to flow outside of the blood vessel. This must be compensated when estimating exact quantitative vessel cross-sectional area from impedance measurements \cite{kassab2004measurement, kassab2005novel, kassab2009novel}. However, when advancing an electrode catheter through a series of veins, Svendsen et al. found that the inverse relationship between measured impedance and cross-sectional area is preserved even when surrounding tissue is present \cite{svendsen2013accurate}. The same has been reported for coronary arteries \cite{kassab2009novel}. Under this assumption, the general morphology of the BN signal should be the same in real tissue, even though the absolute impedance values will be smaller. This would necessitate either more sophisticated normalization techniques or suggest to focus more on signal shape (areas of change) and less on absolute signal value - which we already do by using derivatives in our DTW calculation. The second input of the tracking pipeline, the distance estimation, is expected to suffer from errors in real tissue, specifically systematic underestimation of the traveled distance \cite{maier2023extending}. This could partly be mitigated using simulation data \cite{maier2023extending}, but further mitigation will likely be necessary. Here, fusion of signals from multiple electrode configurations might help, e.g. in limiting the distance error by checking for consistency of estimated distance among multiple configurations, or by constraining the distance estimation to estimates consistent with the locations and BN signal of all tracked electrodes. Replicating the current study with tissue experiments will pose some difficulty, as the aneurysm is a strong feature used to locate the renal arteries and to constrain our localization. Yet most tissue samples and even porcine models for in-vivo experiments will not have aneurysms, and we expect tracking performance to be reduced by the lack of this localization feature.

In this study, we demonstrated that different electrode configurations on the catheter are suitable for distance estimation, BN impedance recording and tracking. In the future, we want to focus on fusion of information from multiple configurations to improve the tracking performance. There is also room for optimization of the number and distribution of electrodes, and regarding the addition of other sensors to compensate for information lacking in the BN approach, like the catheter's shape. We are certain that all of these will benefit from the ability to flexibly produce design iterations using the proposed manufacturing pipeline, making it a basis for future investigation into the topic.

\section{Experimental Section}

\subsection{Preform Preparation for Thermal Drawing}

The PC preform (\text{\O}40\,mm  × 170\,mm height) is printed on an Ultimaker3 (0.05~mm layer thickness, 100\% infill, 270 $^oC$ printing nozzel temperature, 107 $^oC$ bed temperature, 50~mm/s print speed). Thermal drawing used upper/mid/bottom furnace zone temperatures of 120/190/85$^oC$, 1mm/min Downfeed speed, 0.4mm/s initial capstan speed (tuned during the draw by real time monitoring of the laser micrometer).

\subsection{Wire Feeding Platform Preparation} 

Preform and wire feeding platform are fastened to the preform holder bar with bolts, and each wire is threaded from the bobbin, through a guiding hole, and through one of the 16 preform channels using an acupuncture needle (\text{\O} 300$\mu m$). This assembly is then mounted onto the thermal drawing tower.

\subsection{Mechanical Testing Preparation and Execution} 
\label{Section: Mech Test prep}
Five samples (total length 200~mm, gauge length 100~mm) were used per test. 
For \textbf{tensile testing}, the ends of each sample were placed between rubber pads to reduce stress concentration and prevent the jaws of the tensile testing machine from crushing the sample. Rapid-curing epoxy was used to fasten the sample to the pad, increasing adhesion and preventing slip. The  sample is then clamped firmly in between steel jaws and fastened onto the mounts of an Instron 3369 tensile testing machine (Instron, Wycombe, England, UK), using a load cell of 50~kN. Results of force and displacement were recorded at an acquisition rate of 50~Hz by the tester's built in software. 
For \textbf{3-point bending testing}, the catheter is placed on a 2-point supporting stage, with a support span of 50~mm, and depressed using a 3\textsuperscript{rd} point attached to the load cell at a rate of 10~mm/min until failure, or when it has been displaced by 30~mm.
For \textbf{Compression Testing}, the samples were placed in rubber pads, then placed in the jaws of a tensile testing machine, before the sample is compressed by 20~mm.
For \textbf{Torsion Testing}, samples were clamped within two pin vises, with a fixture separation of 100~mm, one of which is fixed to a Dynamixel XL430 servomotor (Robotis, Seoul, South Korea), and the other end fixed to an ATI Gamma force/ torque sensor (ATI industrial automation, North Carolina, US). The catheter is rotated by the servomotor until failure (kinking or fracture).

\subsection{Laser Micromachining Details} \label{Section: Methods-laser}

A linear stage and rotary stage were used to position the catheter under the laser. 
For \textbf{laser profiling}, grooves measured 0.1~mm by 0.8~mm, with 1~mm spacing. We set the 512~nm green laser's power to 10\% via the output attenuator, at 71429~Hz output repetition rate, 15 as hatch number, and 100~mm/s laser speed.
For \textbf{exposing copper wires},  windows were cut of size 1~mm by 0.35~mm. Settings were 7\% output power,  71429~Hz repetition rate, 16 as hatch number, and 100~mm/s speed.
For \textbf{cutting windows out of the heat shrink tube} on top of electrodes, we cut rectangular contours of 0.67~mm by 1~mm. Settings were 1\% output power, 62500~Hz output repetition rate, 64 as layer count, 100~mm/s laser speed. For the copper wire windows, the laser was used to remove the complete window, whereas for the heat shrink tube windows, we only cut the contours of the windows, as to not subject the electrode surface to the laser. Afterwards, the heat shrink windows were removed using a needle.

\subsection{Laser Spot Welding Details}
A Rofin Starweld-XE laser welder (Rofin-Sinar, Michigan, US) was used to weld the $50~\mu m$ copper wires to the inside of the Pt/Ir electrode rings. Settings were empirically optimized to achieve consistent welds. Voltage was 220~V, pulse width 1.2~ms and repetition rate 16~Hz (pulse energy 0.87~J, energy density 480.8~$\frac{J}{{cm}^2}$). It was not necessary to remove the insulation of the copper wire prior to welding, as the welding laser removed it during the process. Before welding, the electrode ring was positioned with its central axis parallel to the ground, then tilted approximately 25\degree\ downwards. The  wire tip was placed at the bottom of the inner surface of the tilted ring, and the laser was activated until a weld visibly formed. 

\subsection{Impedance Measurements}
An Eliko Quadra multifrequency electrical impedance spectroscopy device (Eliko, Tallinn, Estonia) was used to record tetrapolar impedances, 
using a multiplexer front-end with a current excitation setting. The real component of the complex impedance at 1~kHz was used for all our analysis. 

\subsection{FEM Simulations}
Gmsh finite element mesh generator was used for meshing and Elmer FEM as solver, simulating static current flow with Elmer's Static Current Conduction model.
Simulation steps follow \cite{maier2023extending}.
For centerline-based simulations, the vessel centerline is extracted with 3D Slicer’s Vascular Modeling Toolkit and sampled with a control point distance of 0.1~mm. The vessel mesh is decimated and converted to brep via FreeCAD, then imported into Gmsh. The catheter is modeled by extruding a disk along a spline fitted to the centerline.
Stepping through the centerline, one simulation is done at every 10th centerline point (1~mm distance). The catheter tip is modeled symmetrically around this centerline point.

\subsection{Simple Geometry Phantom Preparation}
The phantom was designed in Solidworks (Dassult Systems), then 3D printed using a Bambu Labs printer. It consists of cylindrical bores of increasing diameter ( 10~mm, 9~mm, 8~mm, 7~mm, 6~mm).
We connected flexible tubes to the phantom and filled it with physiological saline solution (0.9\%), representing blood.

\subsection{Silicon Phantom Preparation}
After filling the phantom with 0.9\% saline, we threaded a commercial 0.89~mm guidewire from the left iliac artery (Figure \ref{fig:AAA_and_distance}A, sections S1, S2) through the aneurysm (S3) and into the healthy part of the abdominal aorta (S4, S5). Then, we advanced our prototype over the guidewire. For forward runs, the initial catheter location was such that the most proximal sensing electrode was 2 cm deeper into the phantom than the stationary electrode. For pullbacks, we started with the most distal sensing electrode at the end of the phantom. Before each set of recordings, we adjusted the injected current and amplifier gains to maximize signal strength without causing saturation or overflow of the Eliko’s dynamic range.

\subsection{Registration between Electromagnetic and Phantom Coordinate System}
We registered the EM tracking coordinate system to the phantom coordinate system using fiducials. We chose the 8 corners of the acrylic glass box that encases the phantom as fiducial points, which we located in the EM coordinate system. We annotated the same points on the 3D model of the phantom, and registered them using these point-wise corresponding fiducials in ImFusion Suite (ImFusion GmbH, Germany).

\subsection{Preprocessing of Signals before Dynamic Time Warping}
Initially, the live impedance signal from a tetrapolar configuration on the catheter is a time series of impedance values, growing in length with each new measurement from the impedance spectroscope (approx. 17 times per second when measuring 4 channels). This means DTW would need to match two signals with different physical abscissae: the simulated reference signal's abscissa is in millimeters, because it was simulated on a centerline, and the live signal's is in milliseconds. This causes multiple issues. If the catheter reverses direction, i.e. it moves forward and then gets pulled back, matching will fail as DTW can only calculate monotonous alignments. Yet pullbacks are common in  endovascular interventions, to reposition a device
or navigate into branches. Secondly, catheter motion speed strongly affects  signal shape. If the catheter moves very slowly or fast through a phantom region, signal derivatives will be scaled with the movement speed. If the catheter stops in a region where the phantom diameter changes (reference signal shows lots of change), the live signal plateaus instead. This precludes using derivatives in DTW's distance metric. Thirdly, signal length grows rapidly, requiring DTW to  update multiple times per second in realtime, even if no new information is present in the new samples (e.g. if the catheter didn't move).

To mitigate this, we preprocess the impedance signals using the catheter's distance estimates, which are produced at the same rate as the tetrapolar impedances. Each impedance is associated with an estimated distance from the point where the catheter started moving. First, we use the distance estimates to discard parts of the impedance signal during which the catheter moved non-monotonously. For instance, if the catheter advances 5~cm and then retracts 4~cm, we eliminate all impedances measured during both the 4~cm backward motion and the preceding 4 cm of the forward motion, leaving only the impedance signal corresponding to a net 1 cm of monotonous forward movement. Secondly, we use the distance estimates to resample the remaining impedance signal at 1~mm intervals via piecewise linear interpolation.
Finally, we normalize the impedance signal amplitude by calculating the mean value of the first 5~mm of the resampled signal, and dividing the signal by this value. The normalized impedance signal then starts at a signal value of 1.0, with the abscissa representing 1~mm increments. Since we simulated the reference signals based on 1~mm steps on the centerline, both input signals to DTW are now of the same abscissa. Also, non-monotonous segments have been removed leaving DTW to only match monotonous segments, derivatives are independent of motion speed, and both signals are normalized to start at a value of 1.0.

We prefer normalizing based on the first few millimeters of the impedance signal compared to more common methods like standardization. This is because after the catheter has passed this initial distance, the normalization parameters are fixed and new measurements do not change them (as they would when using e.g. mean and standard deviation of the whole signal). 
DTW is calculated iteratively, allowing us to re-use intermediate results from shorter, earlier matching calculations, which makes it very fast to calculate when the impedance signal only grows by a few values. With any scheme where normalization parameters change dynamically with new signal values, we would have to re-normalize the live signal whenever new values are recorded, forcing DTW to restart from scratch every time the normalization parameters change.

\subsection{Weighting Factor for Derivatives}
\label{Section: Weighting Factor}
The factor weighing derivative error against signal error was calculated as the ratio of mean absolute signal value and mean absolute derivative value of the centerline-based simulated signal.

\subsection{Variance-Based Anchor Regions}
For each index in the catheter's normalized impedance signal, the variance of the signal within a causal window of length W (current index plus W-1 preceding indices) is calculated and normalized by the window's mean. Regions where the standard deviation inside the window exceeds a threshold of 0.025 (i.e. variance exceeds 0.00625) are anchor regions. The threshold was empirically tuned to include most regions except for near-plateaus; rationale for the selected value is given in Supporting Information Section 10.

% Acknowledgements
\medskip
\textbf{Acknowledgements} \par 
A. R., H. M., and J. Z. contributed equally to this work.
The authors would like to thank Florent Seichepine for his training and support while using the laser micro-machining facilities, Wenjing Zheng for her help with fabrication and device design, and Celia Riga M.D., Colin Bicknell M.D. for their feedback on the device design. We would further like to thank PD Dr. J. Nadjiri and Prof. Dr. P. Paprottka for their aid in creating a CT scan of the given vascular phantom. 

Alex Ranne is supported by the UKRI CDT in AI for Healthcare under Grant EP/S023283/1, and the TUM Global Incentive Fund. Heiko Maier is supported by TUM International Graduate School of Science and Engineering (IGSSE). Alex Ranne and Heiko Maier are supported by the ICL-TUM Joint Academy of Doctoral Studies (JADS) program. Jinshi Zhao is supported by Cancer Research UK under the grant EDDPJT-May21/100001.  

% References
\medskip

\textbf{Conflict of Interest} \par 
The authors declare that there are no conflict of interest.
\medskip

\textbf{Supporting Information} \par 
Supporting Information is available from the Wiley Online Library or from the author.
\medskip
\bibliographystyle{unsrt}
\bibliography{refs}

\end{justify}
\end{document}